\documentclass[fleqn,usenatbib]{mnras}

\usepackage{newtxtext,newtxmath}
\usepackage[T1]{fontenc}

\DeclareRobustCommand{\VAN}[3]{#2}
\let\VANthebibliography\thebibliography
\def\thebibliography{\DeclareRobustCommand{\VAN}[3]{##3}\VANthebibliography}

\usepackage{graphicx}	% Including figure files
\usepackage{amsmath}	% Advanced maths commands

\newcommand{\Chandra}{{\it Chandra}}
\newcommand{\NICER}{{\it NICER}}
\newcommand{\NuSTAR}{{\it NuSTAR}}
\newcommand{\RXTE}{{\it RXTE}}
\newcommand{\Suzaku}{{\it Suzaku}}
\newcommand{\XMM}{{\it XMM-Newton}}
\newcommand{\psr}{PSR~J1813$-$1749}
\newcommand{\cxo}{CXOU~J182913.1$-$125113}

\newcommand{\snr}{G18.95$-$1.1}
\newcommand{\psreight}{PSR~J1810$-$0623}
\newcommand{\psrnine}{PSR~J1933$-$6211}
\newcommand{\nudot}{\dot{\nu}}
\newcommand{\NH}{N_{\rm H}}
\newcommand{\fabs}{f^{\rm abs}_{3-79}}
\newcommand{\fabstwo}{f^{\rm abs}_{2-10}}
\newcommand{\fabsthree}{f^{\rm abs}_{0.3-10}}
\newcommand{\funabsthree}{f^{\rm unabs}_{0.3-10}}
\newcommand{\Msun}{M_\odot}

\title[X-ray portraits of four neutron stars]{Portraits of the young and old in X-ray: New millisecond pulsar detections and updated characterization of young neutron stars}

\author[W. C. G. Ho, P. H. Wang and S. Guillot]{
Wynn C. G. Ho,$^{1}$\thanks{E-mail: who@haverford.edu}
Patrick H. Wang$^{1}$
and
Sebastien Guillot$^{2}$
\\
$^{1}$Department of Physics and Astronomy, Haverford College, 370 Lancaster Avenue, Haverford, PA 19041, USA\\
$^{2}$Universit\'e de Toulouse, CNES, CNRS, IRAP, Toulouse, France\\
}

\date{Accepted 2026 August 28. Received 2026 August 19; in original form 2026 July 4}

\pubyear{2026}

\begin{document}
\label{firstpage}
\pagerange{\pageref{firstpage}--\pageref{lastpage}}
\maketitle

\begin{abstract}
We present analyses of new and archival \Chandra\ and \NuSTAR\ data on
two old millisecond pulsars, a young highly-energetic pulsar, and
a young neutron star.
\psreight\ is a recently-discovered 4.55~ms radio pulsar in a binary with
a possible carbon-oxygen white dwarf companion.
We tentatively detect the pulsar for the first time in X-ray using a new short
\Chandra\ HRC-I observation.  Its faintness suggests a distance more
in accord with the further ($\approx 1\mbox{ kpc}$) of its dispersion
measure-inferred distances.
\psrnine\ is a 3.54~ms radio pulsar also with a carbon-oxygen white
dwarf companion.
Our first-time X-ray detection in an archival HRC-S image indicates
its X-ray luminosity is similar to that of the $1.9~\Msun$ pulsar
PSR~J1614$-$2230.
Next, \psr\ is a 1--2~kyr pulsar with one of the highest known spin-down
luminosities.
We measure its spin period of 44.7~ms using 2025 \NuSTAR\ and 2026 \Chandra\
ACIS-S data and update its spin evolution model.
We show for the first time the pulse profile of \psr\ at 3--79~keV, and
we find that the spectra of the pulsar and pulsar wind nebula at these
energies did not change significantly between observations in 2018 and 2025.
Finally, for the 5--10~kyr neutron star \cxo\ in the supernova remnant \snr,
we combine ACIS-I observations from 2009 and 2024 to better measure the
neutron star's spectrum.
\end{abstract}


%%%%%%%%%%%%%%%%%%%%%%%%%%%%%%%%%%%%%%%%%%%%%%%%%%

%%%%%%%%%%%%%%%%% BODY OF PAPER %%%%%%%%%%%%%%%%%%

\section{Introduction} \label{sec:intro}

X-ray observations of neutron stars can provide invaluable information
about their properties.
For example, thermal X-ray emission from hot spots on the surface of
rapidly-rotating neutron stars, or millisecond pulsars, allowed
measurements of the stellar mass and radius and thus constraints on the
equation of state of dense nuclear matter
(e.g., \citealt{dittmannetal24,salmietal24,vinciguerraetal24,mauviardetal25,milleretal26},
for recent works).
Non-thermal X-rays from young neutron stars give insights into the pulsar
magnetosphere and emission geometry and particle acceleration processes
(see, e.g., \citealt{becker09,enotoetal19}, for review).
Regular monitoring of young pulsars, whose spin in some cases are only
detected in X-ray, enabled limits to be placed on stellar ellipticities
via gravitational wave searches (see, e.g., \citealt{abacetal25,abacetal26}).
In this work, we report on the first X-ray detection of two old
millisecond pulsars, \psreight\ and \psrnine, albeit the former weakly.
We present new measurements of the spin of the young pulsar, \psr, and
new and archival spectra of \psr\ and the young neutron star \cxo.
Below we briefly describe each pulsar.

\psreight\ is a 4.55~ms radio pulsar that was recently discovered and
monitored by the Five-hundred-meter Aperture Spherical radio Telescope (FAST)
and Green Bank Telescope, and it is in a 15.4~d orbit with a
$\sim 0.6\,M_\odot$ carbon-oxygen white dwarf \citep{zhangetal26}.
From its radio dispersion measure (DM), two possible distances to \psreight\
were derived, depending on which standard model of Galactic electron density
is used, either 207~pc using YMW16 \citep{yaoetal17} or
1.024~kpc using NE2001 \citep{cordeslazio03}.
We note that the new model NE2025 \citep{ockercordes26} yields a distance
of 274~pc.
If one of the shorter distances is correct,
\psreight\ would be one of the closest millisecond pulsars known and
potentially detectable and bright in X-ray, like a number of other nearby
millisecond pulsars observed by the Neutron star Interior Composition Explorer
(\NICER) to measure their mass and radius
(e.g., \citealt{bogdanovetal19,guillotetal19,wolffetal21}).

\psr\ is a highly energetic pulsar that produces a pulsar wind nebula (PWN) and
is associated with the gamma-ray/TeV source IGR~J18135$-$1751/HESS~J1813$-$178
and supernova remnant G12.82$-$0.02 \citep{helfandetal07}.
The pulsar is young, with an age of 1000--2200~yr
\citep{broganetal05,dzibrodriguez21} and at a distance of 6--14~kpc
\citep{camiloetal21}.
Even though the pulsar is detected in radio,
its radio pulses suffer from extremely high scattering \citep{camiloetal21}.
Meanwhile, its X-ray pulse profile has a broad single peak and high pulsed
fraction below 10~keV
\citep{gotthelfhalpern09,halpernetal12,kuiperhermsen15,hoetal20,takataetal24}.
The spin evolution of \psr\ has been tracked in recent years using \Chandra\
\citep{hoetal22,hoetal26},
and X-ray spectra of the pulsar and PWN have been studied by
\citet{hoetal20,bambaetal22,takataetal24}.

\cxo\ is a young neutron star found by \Chandra\ in \snr\
\citep{tullmannetal10},
which is a 4--10~kyr supernova remnant at a distance of
$\sim 2-4\mbox{ kpc}$
(see \citealt{hollandashfordetal25}, and references therein).
The neutron star's proper motion was recently measured
using two sets of \Chandra\ images taken 15 years apart
\citep{hollandashfordetal25}.
Its X-ray spectrum was studied using the first of these observations
in 2009 and could be fit by an absorbed power law with $\Gamma=1.6$
\citep{tullmannetal10}.

\psrnine\ is a 3.54~ms radio pulsar in a 12.8~d orbit \citep{jacobyetal07}.
The binary companion is a carbon-oxygen white dwarf that has a mass of
$\approx 0.43\,M_\odot$,
and the system's distance of 1.6~kpc is two to three
times greater than that inferred from the pulsar's DM \citep{geyeretal23}.
\Chandra\ observed the field of \psrnine\ for 20~ks in 2007 (ObsID 9115),
and \citet{prinzbecker15} derived an upper limit on the 0.1--2~keV flux of
$2.2\times10^{-14}\mbox{ erg s$^{-1}$ cm$^{-2}$}$,
assuming a hydrogen absorption column density
$\NH=4\times10^{21}\mbox{ cm$^{-2}$}$ and
a power law (or blackbody) spectrum with $\Gamma=1.7$.

An outline of the paper is as follows.
Section~\ref{sec:data} describes the \Chandra\ and \NuSTAR\ data
analysed in this work.
Section~\ref{sec:results} presents our results.
Section~\ref{sec:discuss} summarizes and discusses our findings.

\section{Observations and data analysis} \label{sec:data}

%-------------------------------------------------
\begin{table}
\centering
\caption{Log of \Chandra\ ACIS and HRC and \NuSTAR\ FPM observations
used in this work.
ACIS-S observation conducted in continuous clocking (CC) mode.}
\label{tab:data}
\begin{tabular}{lcccc}
  \hline
Source & Instrument & ObsID & Observation & Exposure \\
PSR/CXOU & & & date & (ks) \\
  \hline
J1810$-$0623 & HRC-I & 32291 & 2026 May 5 & 5 \\
J1813$-$1749 & FPM & 30364003002 & 2018 Mar 25 & 26 \\
& FPM & 31101010002 & 2025 Oct 17 & 21 \\
& ACIS-S (CC) & 31125 & 2026 Mar 16 & 20 \\
J1829$-$1251 & ACIS-I & 10098 & 2009 Aug 3 & 45 \\
& ACIS-I & 26656 & 2024 Jul 11 & 21 \\
& ACIS-I & 29478 & 2024 Jul 11 & 23 \\
J1933$-$6211 & HRC-S & 9115 & 2007 Dec 9 & 20 \\
\hline
\end{tabular}
\end{table}
%-------------------------------------------------

\subsection{\Chandra\ ACIS and HRC data} \label{sec:chandra}

\Chandra\ observed each of our four sources with either
ACIS-I (for \cxo), ACIS-S in continuous clocking (CC) mode (for \psr),
HRC-I (for \psreight), or HRC-S (for \psrnine).
A log of these observations is given in Table~\ref{tab:data}.
We reprocess all data using \texttt{chandra\_repro} with default settings
from the Chandra Interactive Analysis of Observations (CIAO) package
version 4.18 and Calibration Database (CALDB) 4.12.3 \citep{fruscioneetal06}.

For timing analysis of \psr\ with CC mode data,
we follow the recommended procedure\footnote{\url{https://cxc.cfa.harvard.edu/ciao/caveats/acis_cc_mode.html}}
and use \texttt{specextract} to extract source events from the
one-dimensional image along a 1.5~arcsec length centred on the
pulsar position.
We select only events in the 2--8~keV energy range.
We transform the selected event time stamps from Terrestrial Time (TT)
to Barycentric Dynamical Time (TDB) using \texttt{axbary} and the pulsar
position.
Acceleration searches for the spin frequency are conducted using
PRESTO \citep{ransometal02} with a time bin of 2.85~ms.
Data are folded at the candidate pulse frequency using \texttt{prepfold},
and a refined frequency is determined.

For spectral analysis of \cxo,
source counts are extracted from a 5~arcsec radius circular region
(a 5.2~arcsec radius is used in \citealt{tullmannetal10}),
and an annulus with inner and outer radii of 10 and 20~arcsec is used
for background determination.
We combine the two 2024 spectra using \texttt{combine\_spectra} and
bin separately the 2009 spectrum and the combined 2024 spectrum using
\texttt{dmgroup} until a signal-to-noise ratio of 3 is achieved
(same as in \citealt{tullmannetal10}).

HRC does not possess the capabilities needed for detailed spectral analyses
of \psreight\ and \psrnine.
Thus we simply use \texttt{ds9} to extract photons within specified regions
(see below) in order to determine source and background counts and count rates.

\subsection{\NuSTAR\ data} \label{sec:nustar}

\NuSTAR\ observed \psr\ twice, in 2018 and last year on 2025 October 17
(see Table~\ref{tab:data}).
We reprocess these data using \texttt{nupipeline} and \texttt{nuproducts}
of HEASoft~6.35.2 \citep{heasarc14} and NUSTARDAS~2025-03-11\_v2.1.5
and barycentre events with the pulsar's position and clock corrections
20100101v211 for the 2018 data and 20100101v228 for the 2025 data.
The source extraction region is a circle of 80~arcsec radius,
and the background region is a circle of
the same size and placed in a field with low counts.
For spectral analyses, we bin with a minimum of 100 counts per bin.
For timing analysis, we use \texttt{xselect} to choose only events in
the 3--50~keV range.
We use PRESTO and \texttt{prepfold} to perform pulsation searches on
the merged FPMA and FPMB event list and to determine the spin frequency.

To extract rotation phase-resolved pulsed spectra,
we first use \texttt{photonphase} in PINT \citep{luoetal21} and the spin
frequency to determine the rotation phase of each photon in the data.
We then use \texttt{maketime} to determine good time intervals (GTIs) for
on-pulse emission, i.e, photons with a rotation phase around the peak
of the pulse profile, and off-pulse emission,
which are photons with a phase outside the on-pulse phase range.
Finally, we run \texttt{nuproducts} with these GTIs to produce on and
off-pulse spectra.
The off-pulse spectra can be used as the background for the on-pulse spectra,
such that the final pulsed spectra are the result of on minus off-pulse
emissions.

\subsection{Spectral modelling} \label{sec:spectralmodel}

We perform spectral fitting using Xspec 12.15.0d \citep{arnaud96}.
The spectral models used here are composed of two components.
The first component accounts for photoelectric absorption
by the interstellar medium, i.e., \texttt{tbabs} with abundances from
\citet{wilmsetal00} and cross-sections from \citet{verneretal96} and
parametrized by the effective hydrogen column density $\NH$.
A second component, which models the intrinsic spectrum of the source,
is a powerlaw (PL; \texttt{powerlaw}).
Once a best-fit is found, we use \texttt{cflux} to calculate flux and its
uncertainties while holding the powerlaw normalization at its best-fitting
value.

\section{Results} \label{sec:results}

We present first our results for the two old binary millisecond pulsar
systems, \psreight\ and \psrnine, which only have HRC imaging data, and
then results for the two young sources with spectral and/or timing data,
\psr\ and \cxo.

\subsection{\psreight} \label{sec:psrj1810}

\begin{figure}
\begin{center}
\includegraphics[width=0.9\columnwidth]{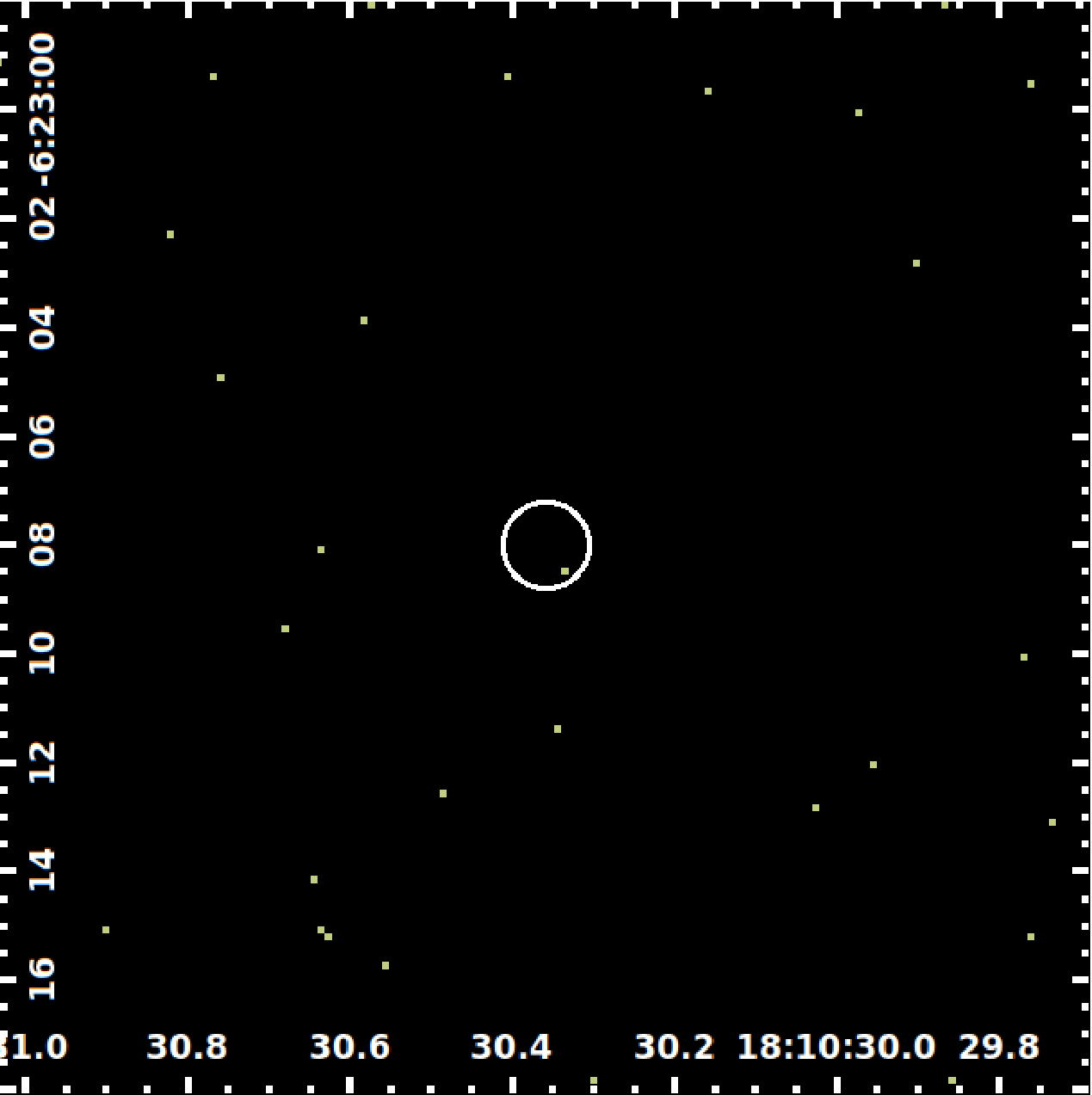}
\caption{20~arcsec~$\times$~20~arcsec field of \psreight\ from a 5~ks
\Chandra\ HRC-I observation (ObsID 32291).
Circle with 0.8~arcsec radius is centred on \psreight.
\label{fig:psrj1810}}
\end{center}
\end{figure}

Figure~\ref{fig:psrj1810} shows the 5~ks HRC-I image of the field of \psreight,
centred on the pulsar position from \citet{zhangetal26} of
R.A.$=18^{\rm h}10^{\rm m}30.3585^{\rm s}\pm0.0002^{\rm s}$,
decl.$=-06^\circ23\arcmin08.015\arcsec\pm0.009\arcsec$.
The image shows 1~photon detected 0.6~arcsec from the radio-determined
position, which is known to an accuracy of about 0.01~arcsec.
This can be compared to the 0.4~arcsec absolute astrometric accuracy of HRC
and the point spread function (PSF) of about 0.8~arcsec for an enclosed
counts fraction (ECF) of 85~percent.
We estimate a background of 0.1~counts in any 0.6~arcsec radius circle,
based on 1023 photons detected within 60~arcsec of the source.
Therefore, there is a 10~percent probability for any 0.6~arcsec radius circle
to enclose a single photon by chance,
and we associate the detected photon with the pulsar at 90~percent confidence.
This then tentatively yields a HRC-I count rate of $0.0002\mbox{ s$^{-1}$}$
for \psreight.

%-------------------------------------------------
%-------------------------------------------------

Due to the lack of information that can be obtained from the one count
detected in the HRC-I observation, we use PIMMS to estimate the X-ray
flux of \psreight.
The total Galactic contribution to the absorption column density in the
direction of \psreight\ is
$\NH=3.0\times10^{21}\mbox{ cm$^{-2}$}$ \citep{hi4pi16}.
In contrast, a value of $7.3\times10^{20}\mbox{ cm$^{-2}$}$ is inferred
using the relation of \citet{heetal13} between $\NH$ and radio
dispersion measure, and this lower value may be more accurate if the pulsar
is as close as 200~pc (see Section~\ref{sec:intro}).
For the intrinsic X-ray spectrum of \psreight, we assume it has the same
properties as those of PSR~J1614$-$2230,
which is the only other pulsar of a similar type to \psreight\ (and \psrnine)
to be detected in X-ray (see Section~\ref{sec:discuss}).
In particular, the spectrum of PSR~J1614$-$2230 has been measured using
\Chandra\ \citep{marellietal11} and \XMM\ \citep{pancrazietal12,wolffetal21}
and fit using an absorbed blackbody, an absorbed double blackbody,
or an absorbed blackbody plus powerlaw model.
With a HRC-I count rate of $0.0002\mbox{ s$^{-1}$}$, the two possible
values for $\NH$ given above, and best-fitting spectral model parameters
from \citet{pancrazietal12,wolffetal21}, we estimate a 0.3--10~keV
absorbed flux $\fabsthree=(4\pm2)\times10^{-15}\mbox{ erg s$^{-1}$ cm$^{-2}$}$
and unabsorbed flux
$\funabsthree=(6\pm3)\times10^{-15}\mbox{ erg s$^{-1}$ cm$^{-2}$}$.
The implication on the distance to \psreight\ of the estimated X-ray flux,
compared to the radio DM, is discussed in Section~\ref{sec:discuss}.

\subsection{\psrnine} \label{sec:psrj1933}

\begin{figure}
\begin{center}
\includegraphics[width=0.9\columnwidth]{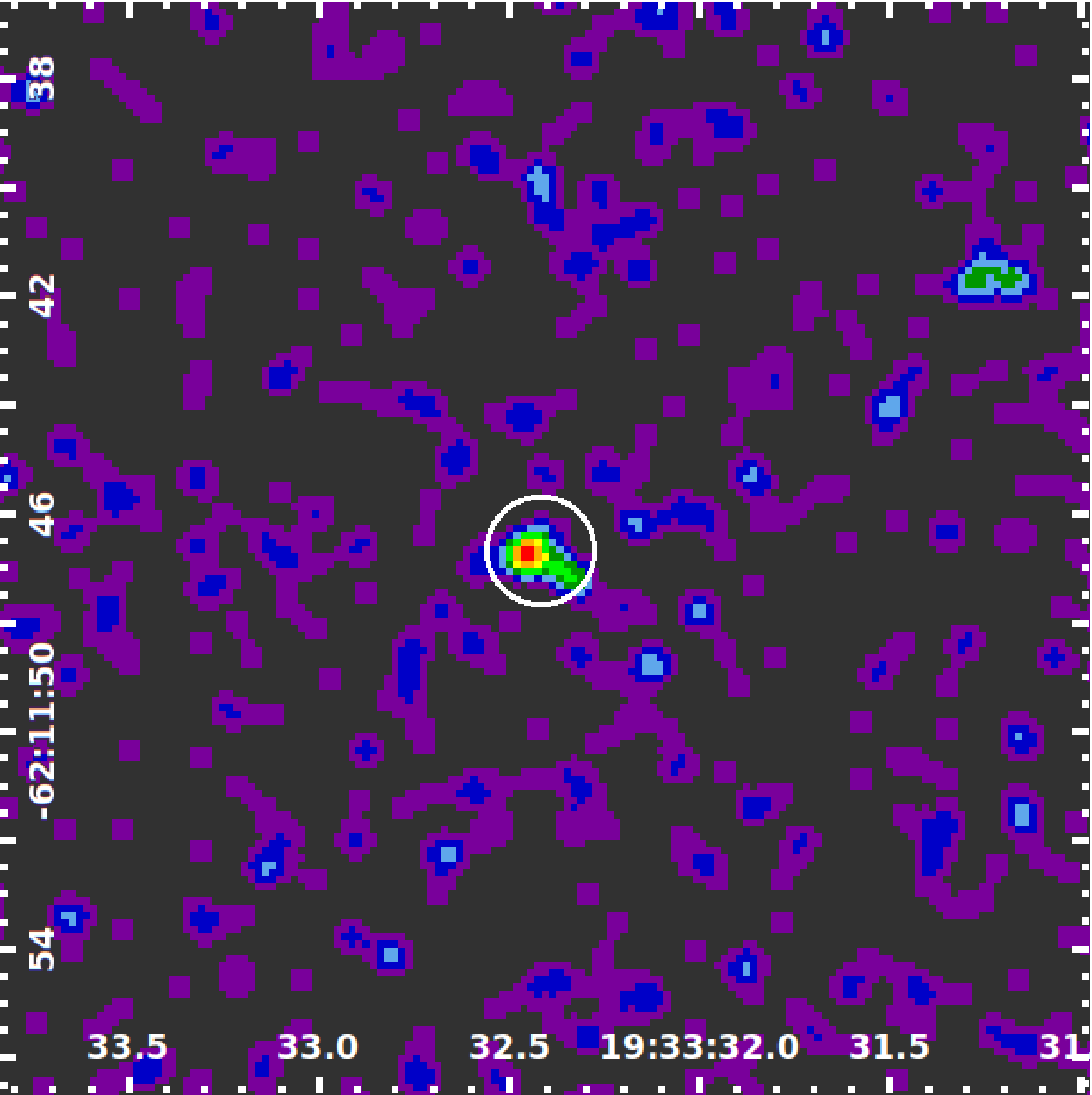}
\caption{20~arcsec~$\times$~20~arcsec field of \psrnine\ from a 20~ks
\Chandra\ HRC-S observation (ObsID 9115).
Image is smoothed using a Gaussian kernel with a 4~pixel radius.
Circle with 1~arcsec radius is centred on \psrnine.
\label{fig:psrj1933}}
\end{center}
\end{figure}

Figure~\ref{fig:psrj1933} shows the 20~ks HRC-S image of the field of \psrnine,
centred on the pulsar position from \citet{geyeretal23} of
R.A.$=19^{\rm h}33^{\rm m}32.413992^{\rm s}\pm0.000009^{\rm s}$,
decl.$=-62^\circ11\arcmin46.70233\arcsec\pm0.00009\arcsec$, with 1$\sigma$
uncertainties.
17 photons are detected within a 1~arcsec radius circle around the pulsar,
and 4 of these are expected to be background events, as determined from
14383 photons detected within a 60~arcsec radius circle.
Thus we estimate the number of counts from \psr\ to be $13\pm4$ in 20~ks
and a HRC-S count rate of $0.00065\pm0.0002\mbox{ s$^{-1}$}$.

%-------------------------------------------------
%-------------------------------------------------

As in the case of \psreight\ above, we use PIMMS to estimate the X-ray flux
of \psrnine.
The total absorption column density $\NH$ in the direction of \psrnine\ is
$5.1\times10^{20}\mbox{ cm$^{-2}$}$ \citep{hi4pi16}, while the
column density determined from the $\NH$--DM relation of \citet{heetal13}
is $3.5\times10^{20}\mbox{ cm$^{-2}$}$.
The two values are not significantly different even though the source
distance is only $1.6^{+0.2}_{-0.3}\mbox{ kpc}$ \citep{geyeretal23},
possibly because the pulsar is 29~degrees out of the Galactic plane.
We consider both values in our estimate.
With a HRC-S count rate of $0.00065\pm0.0002\mbox{ s$^{-1}$}$,
two possible values for $\NH$,
and spectral model parameters of PSR~J1614$-$2230,
we determine a 0.3--10~keV absorbed flux
$\fabsthree=(8\pm4)\times10^{-15}\mbox{ erg s$^{-1}$ cm$^{-2}$}$
and unabsorbed flux
$\funabsthree=(9\pm4)\times10^{-15}\mbox{ erg s$^{-1}$ cm$^{-2}$}$
for \psrnine.
The latter is effectively what we obtain by placing the
$\sim3\times10^{30}\mbox{ erg s$^{-1}$}$ luminosity of PSR~J1614$-$2230
at 1.6~kpc.

\subsection{\psr} \label{sec:psrj1813}

\begin{figure}
\includegraphics[width=\columnwidth]{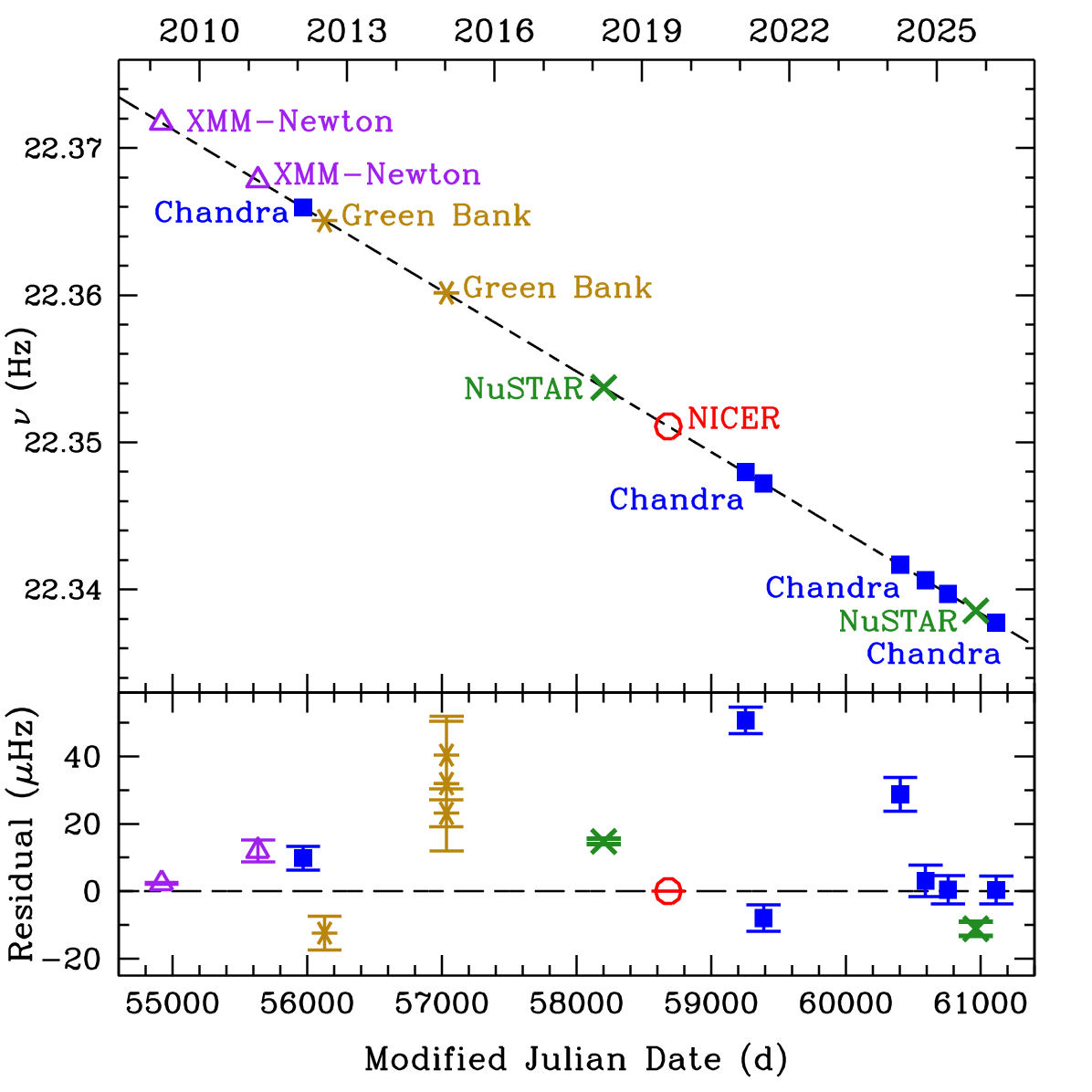}
\caption{Spin frequency $\nu$ of \psr\ (top) and difference between
best-fit linear model and data (bottom) as functions of time.
Errors are 1$\sigma$ uncertainty.
Measurements of $\nu$ are made using \XMM\ (triangles), \Chandra\ (squares),
Green Bank Telescope in radio (stars), \NuSTAR\ (crosses), and \NICER\ (circle).
Dashed line shows a linear fit of all $\nu$ measurements with
best-fit $\nudot=-6.3443\times10^{-11}\mbox{ Hz s$^{-1}$}$.
\label{fig:psrj1813spin}}
\end{figure}

Our timing analysis of the 2025 \NuSTAR\ and 2026 \Chandra\ data
(see Table~\ref{tab:data} and Section~\ref{sec:data}) yields new
measurements of the spin frequency of \psr.
We find $\nu=22.3385512(21)\mbox{ Hz}$ on MJD~60965.37 and
$\nu=22.3377380(41)\mbox{ Hz}$ on MJD~61115.82,
where the $1\sigma$ uncertainty in the last digit is given in parentheses.
These measurements are shown in Figure~\ref{fig:psrj1813spin},
as well as those obtained previously using \Chandra\ and \XMM\
\citep{halpernetal12,hoetal22,hoetal26},
Green Bank Telescope at radio wavelengths \citep{camiloetal21},
\NICER\ \citep{hoetal20}, and \NuSTAR\ \citep{hoetal22}.
Fitting a simple linear decline in spin frequency yields a best-fit
spin-down rate $\nudot=(-6.34427\pm0.00060)\times10^{-11}\mbox{ Hz s$^{-1}$}$.
Table~\ref{tab:psrj1813} presents the resulting 17~year timing model.
Residuals from the timing model,
as seen in the bottom panel of Figure~\ref{fig:psrj1813spin},
suggest \psr\ may undergo glitches with sizes as large as a
few tens of $\mu$Hz, which are typical for young pulsars.
Unfortunately, the irregular cadence of observations thus far does not permit
more definitive conclusions about possible glitches or their properties.

%-------------------------------------------------
%-------------------------------------------------

%-------------------------------------------------
\begin{table}
\centering
\caption{Incoherent timing parameters of \psr.
Numbers in parentheses are 1$\sigma$ uncertainty in last digit.
The position is from VLA data (MJD 58119), with uncertainties of
$\sim0\fs009$ and $\sim0\farcs13$, and
proper motion has uncertainties of 3.7 and $6.7\mbox{ mas yr$^{-1}$}$ in
$\mu_\alpha\cos\delta$ and $\mu_\delta$, respectively \citep{dzibrodriguez21}.}
\label{tab:psrj1813}
\begin{tabular}{lc}
\hline
Parameter & Value \\
\hline
R.A. $\alpha$ (J2000) & $18^{\rm h}13^{\rm m}35\fs173$ \\
Decl. $\delta$ (J2000) & $-17\degr49\arcmin57\farcs75$ \\
Solar system ephemeris & DE405 \\
Range of dates (MJD) & 54918.14$-$61115.82 \\
Epoch (MJD TDB) & 58681.04 \\
Frequency $\nu$ (Hz) & 22.3510838(2) \\
Freq.\ 1st derivative $\nudot$ (Hz s$^{-1}$) & $-6.3443(6)\times10^{-11}$ \\
Proper motion $\mu_\alpha\cos\delta$ (mas yr$^{-1}$) & $-5.0$ \\
Proper motion $\mu_\delta$ (mas yr$^{-1}$) & $-13.2$ \\
\hline
\end{tabular}
\end{table}
%-------------------------------------------------

There have been two previous studies of the 2018 \NuSTAR\ spectra of
\psr, one of the combined PWN+pulsar and one of the pulsar itself.
\citet{bambaetal22} examined the total PWN+pulsar spectrum,
supplementing \NuSTAR\ data with \Suzaku\ data at 2--10~keV.
They were able to fit the combined spectra using an absorbed powerlaw,
broken powerlaw, or cut-off powerlaw model.
For an absorbed powerlaw, they found
$\NH=(1.01\pm0.03)\times10^{23}\mbox{ cm$^{-2}$}$,
$\Gamma=2.11^{+0.04}_{-0.03}$, and absorbed
2--10~keV flux $\fabstwo=7.77\times10^{-12}\mbox{ erg s$^{-1}$ cm$^{-2}$}$,
with uncertainties at 90~percent confidence level.
They also indicated that broken and cut-off powerlaw models are better fits.
Meanwhile, \citet{takataetal24} extracted the pulsed spectrum of \psr\
from the same 2018 \NuSTAR\ data,
as well as the pulsed spectra at 0.3--10~keV from \NICER\ and \XMM\ pn data.
They modeled the combined spectra using an absorbed powerlaw and
found a best-fit with $\NH=(1.3\pm0.4)\times10^{23}\mbox{ cm$^{-2}$}$,
$\Gamma=1.8\pm0.3$, and unabsorbed 0.3--150~keV flux of
$(1.1\pm0.4)\times10^{-11}\mbox{ erg s$^{-1}$ cm$^{-2}$}$.

%-------------------------------------------------
\begin{figure}
\includegraphics[width=\columnwidth]{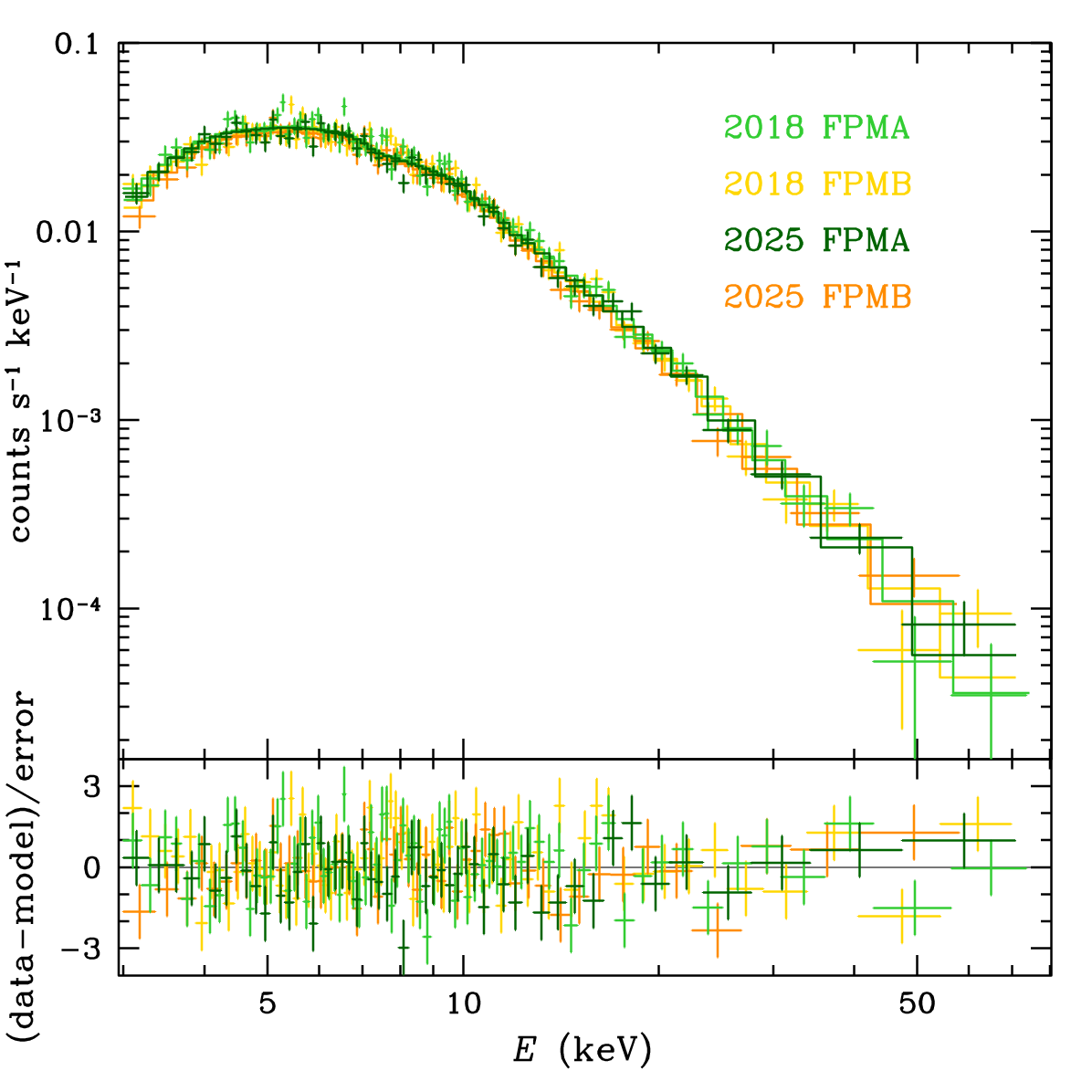}
\caption{
Spectra of \psr\ (PWN+pulsar) from \NuSTAR\ FPMA and FPMB data
taken in 2018 and 2025.
Top panel shows data with 1$\sigma$ errors (crosses) and
spectral model made up of an absorbed powerlaw.
Bottom panel shows $\chi=\mbox{(data-model)/error}$.
}
\label{fig:pwn}
\end{figure}
%-------------------------------------------------

%-------------------------------------------------
\begin{table}
\centering
\caption{Best-fit results of an absorbed powerlaw model to \NuSTAR\
spectra of \psr.
Absorption $\NH$ is in $10^{23}\mbox{ cm$^{-2}$}$
and absorbed 2--10~keV flux $\fabstwo$ and 3--79~keV flux $\fabs$ are in
$10^{-12}\mbox{ erg s$^{-1}$ cm$^{-2}$}$.
Errors are 1$\sigma$, and parameter value without an error is fixed.}
\label{tab:psrj1813sp}
\begin{tabular}{cccccc}
\hline
Year & $\NH$ & $\Gamma$ & $\fabstwo$ & $\fabs$ & $\chi^2$/dof \\
\hline
\multicolumn{6}{c}{PWN+pulsar} \\
2018 & 1.3$\pm$0.1 & 1.92$\pm$0.03 & 8.8$\pm$0.1 & 31.0$\pm$0.7 & 193/146 \\
2025 & 1.4$\pm$0.1 & 1.94$\pm$0.04 & 8.4$\pm$0.1 & 29.4$\pm$0.8 & 85.6/105 \\
2018+2025 & 1.35$\pm$0.08 & 1.93$\pm$0.02 & 8.60$\pm$0.08 & 30.3$\pm$0.5 & 288/254 \\
\\
\multicolumn{6}{c}{pulsar} \\
2018+2025 & 1.35 & 1.6$\pm$0.3 & 0.58$\pm$0.14 & 3.3$^{+1.3}_{-1.0}$ & 187/188 \\
\hline
\end{tabular}
\end{table}
%-------------------------------------------------

With the 2025 \NuSTAR\ observation of \psr, we are able to perform
new analyses of the 3--79~keV spectra of the PWN and pulsar, improving
upon the work of \cite{bambaetal22,takataetal24}.
Following the procedure described in Section~\ref{sec:nustar},
we extract the PWN+pulsar spectra from the FPMA and FPMB data
and fit the 2018 and 2025
data independently, as well as jointly, with the absorbed powerlaw
model described in Section~\ref{sec:spectralmodel}.
The best-fitting results are shown in Figure~\ref{fig:pwn} and
Table~\ref{tab:psrj1813sp}.
Comparisons of the derived 2025 and 2018 fluxes may indicate a small decrease
in flux between the seven years but only at a significance level of about
$2\sigma$.
Also, even though we only analyse \NuSTAR\ data and thus are restricted to
$\ge3\mbox{ keV}$, we obtain $\NH=(1.35\pm0.08)\times10^{23}\mbox{ cm$^{-2}$}$,
which is higher than that found by \citet{bambaetal22} whose spectra
extend down to 2~keV.
On the other hand, our value agrees with that of \citet{takataetal24}
(see also \citealt{hoetal20}), whose spectra extend even further down to
0.3~keV, where $\NH$ can be better constrained.
Finally, fits to our spectra using a broken powerlaw or cut-off powerlaw
did not yield noticeably improved results.

\begin{figure}
\includegraphics[width=\columnwidth]{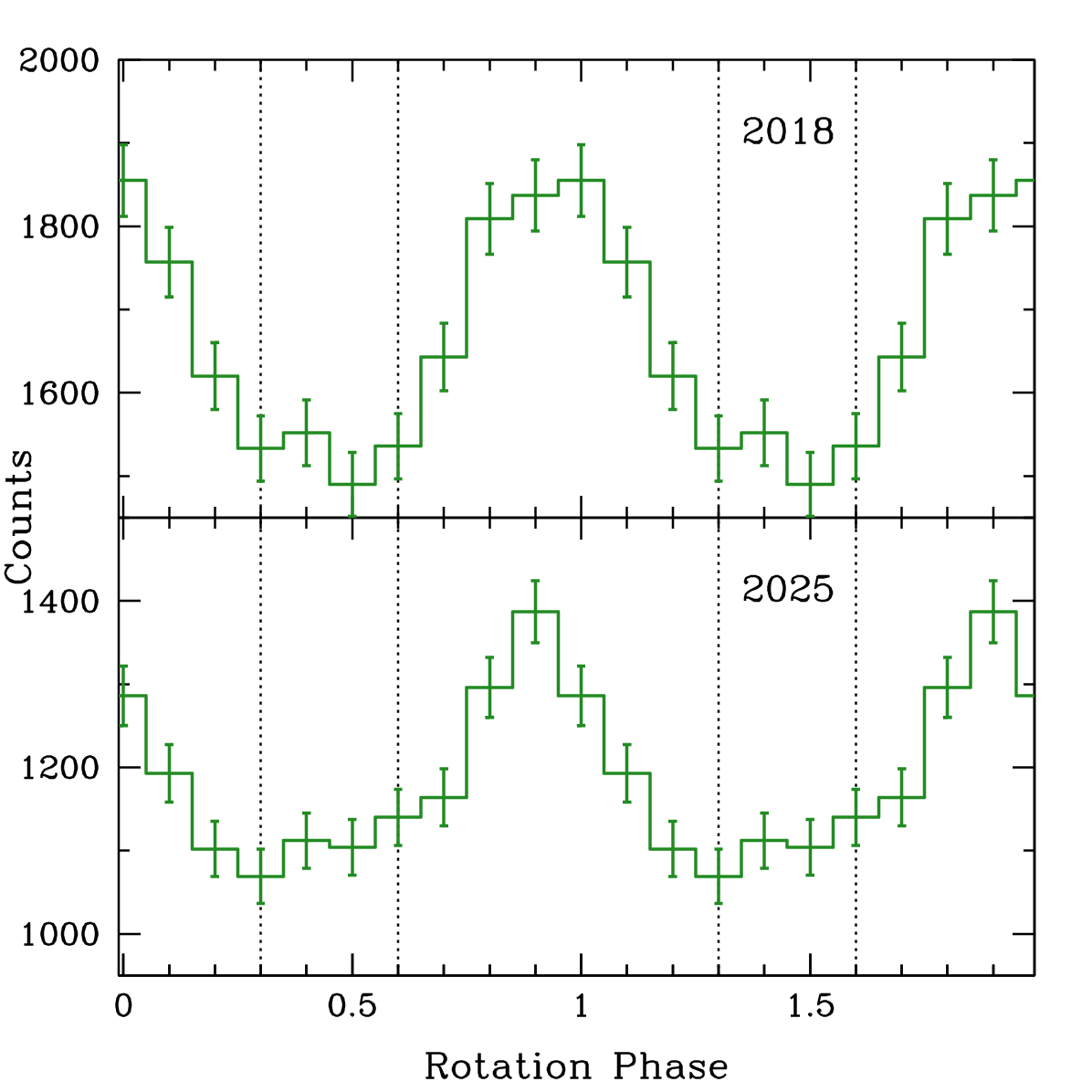}
\caption{
Pulse profile of \psr\ at 3--79~keV from \NuSTAR\ data,
26~ks in 2018 (top) and 21~ks in 2025 (bottom).
Errors are 1$\sigma$ uncertainty.
Two rotation cycles are shown for clarity.
Phase alignment between the two epochs is arbitrary.
Dotted lines indicate phase ranges chosen for on-pulse (phase~$<0.3$ or $>0.6$)
and off-pulse ($0.3<\mbox{phase}<0.6$) spectra.
}
\label{fig:psrj1813pp}
\end{figure}

To extract pulsed spectra of \psr, i.e., emission that is intrinsic
to the pulsar itself, we determine peak on-pulse emission at rotation
phases below 0.3 or above 0.6 and off-pulse emission at phases outside
the on-pulse ranges (see Figure~\ref{fig:psrj1813pp}).
The off-pulse spectra are used as the background for the on-pulse spectra,
such that the final pulsed spectra are the result of on minus off.
As seen from Figure~\ref{fig:psrj1813pp}, the 3--79~keV pulse profile is
broad and single peaked, similar to the pulse profile at lower energies
(see, e.g., \citealt{halpernetal12} for the \Chandra\ pulse profile and
\citealt{kuiperhermsen15} for the \RXTE\ pulse profile at 8--27~keV)
but has a pulsed fraction of $\sim0.12$ (also see next).
A Kolmogorov-Smirnov (KS) test indicates no significant difference between
the normalized pulse profiles from 2018 and 2025.

%-------------------------------------------------
%-------------------------------------------------

\begin{figure}
\includegraphics[width=\columnwidth]{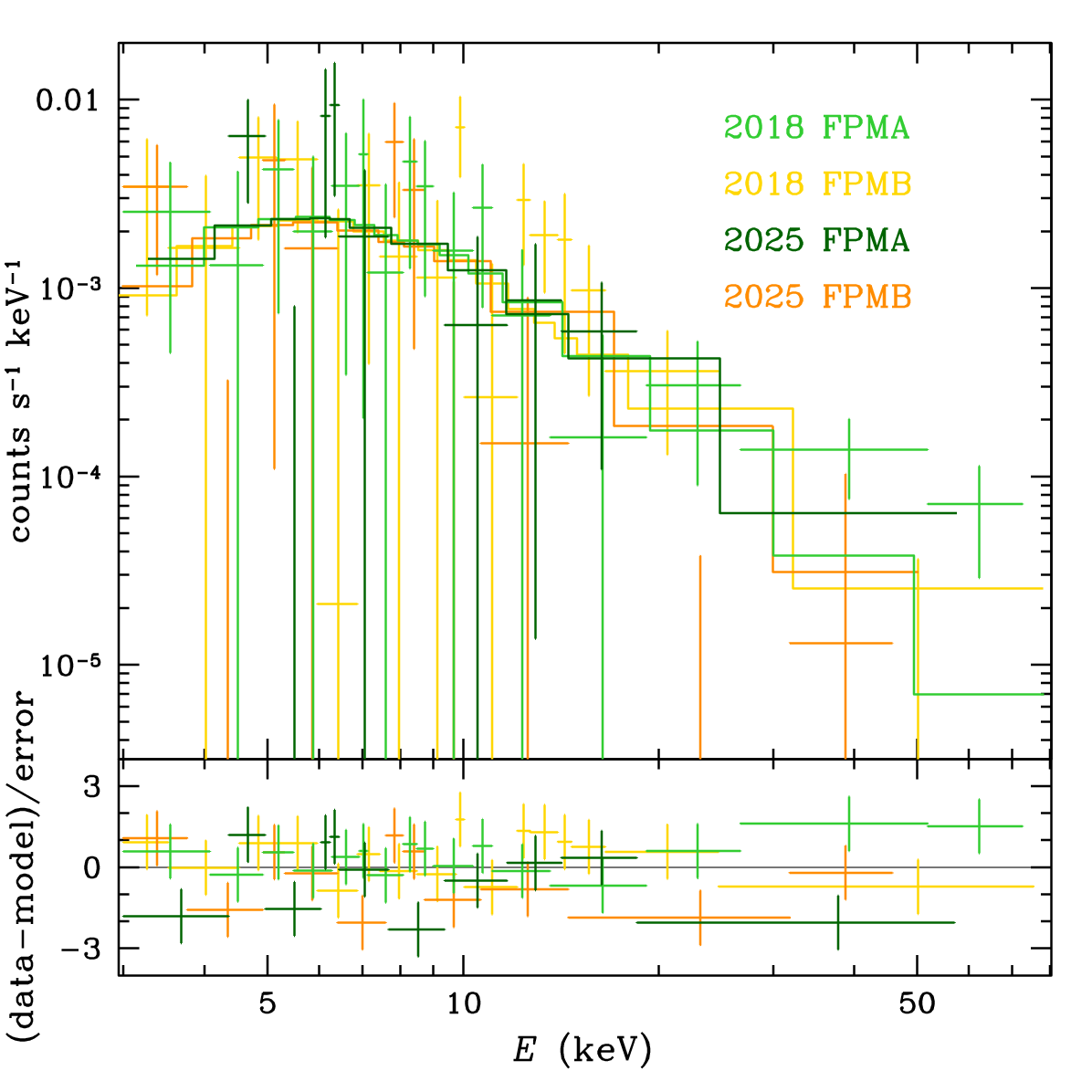}
\caption{
Pulsed spectra of \psr\ from \NuSTAR\ FPMA and FPMB data taken in 2018 and 2025.
Top panel shows data with 1$\sigma$ errors (crosses) and
spectral model made up of an absorbed powerlaw.
Bottom panel shows $\chi=\mbox{(data-model)/error}$.
}
\label{fig:psrj1813sp}
\end{figure}

Like for the PWN+pulsar spectral analysis, we model the pulsed spectra
using an absorbed powerlaw.
Because of the relatively weak pulsed flux, we combine the 2018 and 2025
spectra and fix the absorption to the value derived from the PWN+pulsar
spectra, $\NH=1.35\times10^{23}\mbox{ cm$^{-2}$}$.
Our best-fitting results are shown in Figure~\ref{fig:psrj1813sp} and
Table~\ref{tab:psrj1813sp}
and are consistent with those found by \citet{takataetal24}.
Comparing the pulsed flux to the PWN+pulsar flux, we estimate an
intrinsic pulsed fraction of $0.11^{+0.04}_{-0.03}$ at 3--79~keV.

%-------------------------------------------------
%-------------------------------------------------

\subsection{\cxo} \label{sec:cxoj1829}

\begin{figure}
\includegraphics[width=\columnwidth]{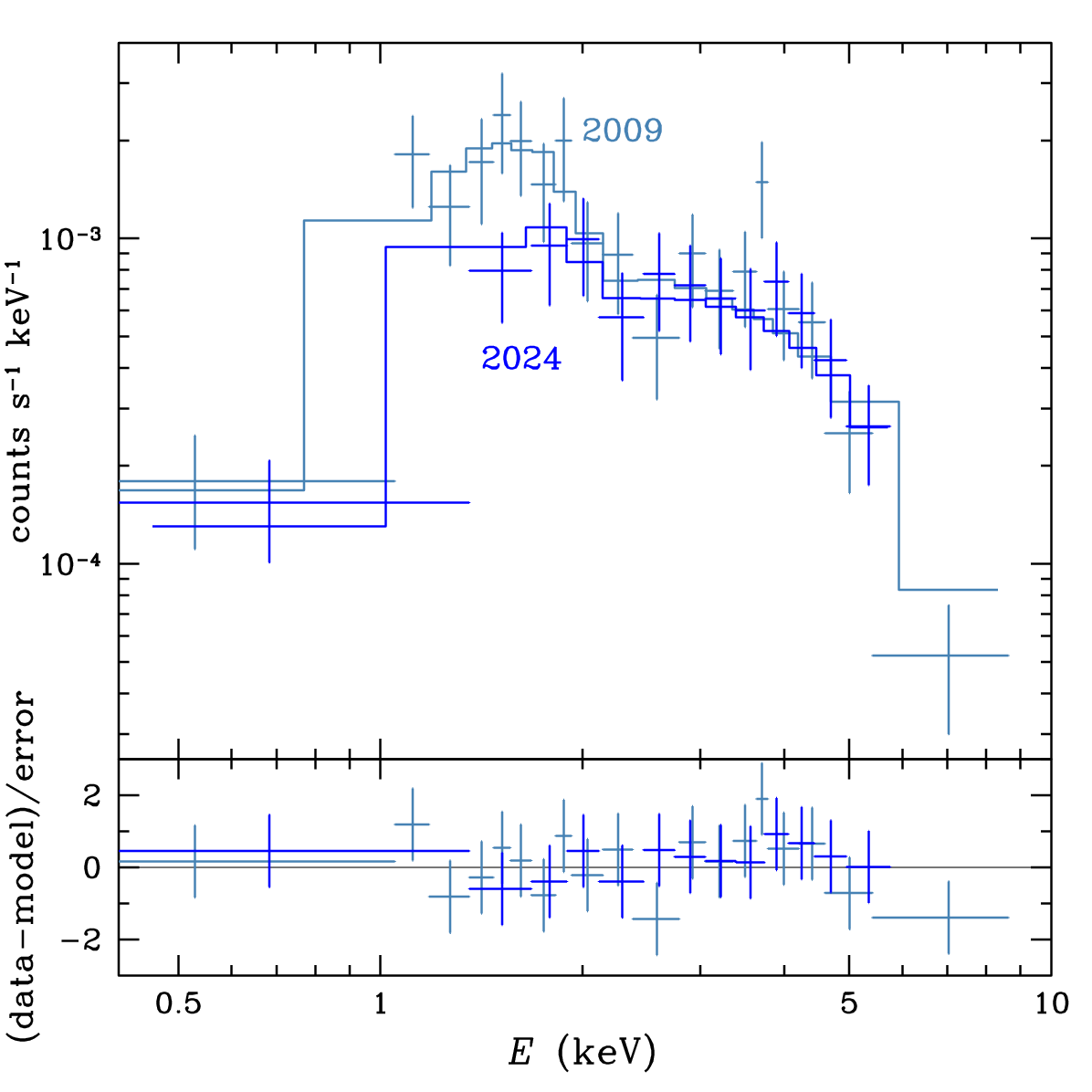}
\caption{
Spectra of \cxo\ from \Chandra\ ACIS-I data from 2009 and 2024.
Top panel shows data with 1$\sigma$ errors (crosses) and
spectral model made up of an absorbed powerlaw.
Bottom panel shows $\chi=\mbox{(data-model)/error}$.
}
\label{fig:cxouj1829}
\end{figure}

Following the procedure described in Section~\ref{sec:chandra},
we extract the spectrum of \cxo\ from the 2009 and 2024 ACIS-I data,
merging the two from 2024,
and fit the two epochs independently, as well as jointly,
with the absorbed powerlaw model described in Section~\ref{sec:spectralmodel}.
The best-fitting results are shown in Figure~\ref{fig:cxouj1829} and
Table~\ref{tab:cxouj1829}.
From the independent fits of the 2009 and 2024 spectra, the flux appears
consistent within errors.
Thus the most reliable results are those from the joint 2009+2024 fit.
We find that \cxo\ is quite absorbed, with a $\NH$ that is consistent
with the $\NH=0.95^{+0.65}_{-0.47}\times10^{22}\mbox{ cm$^{-2}$}$
(at 90~percent confidence level) found using only the 2009 observation by
\citet{tullmannetal10}, as well as the $\NH\sim1\times10^{22}\mbox{ cm$^{-2}$}$
for the supernova remnant \citep{harrusetal04,tullmannetal10,bykovetal22}.
The powerlaw index also agrees with the $\Gamma=1.6\pm0.4$ from
\citet{tullmannetal10}, and our unabsorbed 0.3--10~keV flux
$\funabsthree(1.1\pm0.1)\times10^{-13}\mbox{ erg s$^{-1}$ cm$^{-2}$}$
is consistent with their result.
Finally, we note for completeness that a good fit to the joint spectra is
obtained using an absorbed blackbody.  However, this result is disfavoured
because the best-fit $\NH\sim7\times10^{20}\mbox{ cm$^{-2}$}$ is very low
and the temperature $kT\sim1.0\mbox{ keV}$ is quite high.
Meanwhile, the quality of the data prevents meaningful
constraints using an absorbed blackbody plus powerlaw model.

%-------------------------------------------------
%-------------------------------------------------

%-------------------------------------------------
\begin{table}
\renewcommand{\arraystretch}{1.3}
\centering
\caption{Best-fit results of an absorbed powerlaw model to \Chandra\
spectra of \cxo.
Absorption $\NH$ is in $10^{22}\mbox{ cm$^{-2}$}$
and absorbed 0.3--10~keV flux $\fabsthree$ and 2--10~keV flux $\fabstwo$ are in
$10^{-14}\mbox{ erg s$^{-1}$ cm$^{-2}$}$.
Errors are 1$\sigma$.}
\label{tab:cxouj1829}
\begin{tabular}{cccccc}
\hline
Year & $\NH$ & $\Gamma$ & $\fabsthree$ & $\fabstwo$ & $\chi^2$/dof \\
\hline
2009 & 1.1$^{+0.4}_{-0.3}$ & 1.7$\pm$0.3 & 6.9$\pm$0.9 & 6.0$\pm$0.9 & 13.6/16 \\
2024 & 0.9$^{+1.1}_{-0.8}$ & 1.4$^{+0.6}_{-0.5}$ & 8.6$^{+1.9}_{-1.5}$ & 7.7$^{+1.7}_{-1.5}$ & 2.07/10 \\
2009+2024 & 1.1$^{+0.4}_{-0.3}$ & 1.6$\pm$0.2 & 7.5$^{+0.8}_{-0.7}$ & 6.5$^{+0.8}_{-0.7}$ & 16.8/29 \\
\hline
\end{tabular}
\end{table}
%-------------------------------------------------

\section{Discussion} \label{sec:discuss}

We presented analyses of new \Chandra\ and \NuSTAR\ data,
supplemented by archival data when available, of four neutron stars.
In the case of \psr, we measured its most recent spin frequencies and
updated its timing model.  The spin evolution can be used to search
for possible gravitational waves that are emitted by this young and
highly energetic pulsar.
In fact, timing models such as the one calculated here enabled previous
gravitational wave searches of \psr\ to constrain its fractional stellar
ellipticity to $\lesssim\mbox{a few}\times10^{-5}$
\citep{abbottetal17,abbottetal19,abbottetal22a,abbottetal22b,abacetal25,abacetal26}.
A monitoring campaign at radio wavelengths using the Green Bank Telescope
could allow a rotation-phase coherent timing model to be determined, which
would be extremely useful not only for increasing the sensitivity of
gravitational wave searches but to also see if \psr\ undergoes large and
frequent glitches like the Vela pulsar and PSR~J0537$-$6910.
We measured for the first time the pulse profile of \psr\ up to 79~keV thanks
to the new \NuSTAR\ data, which was previously only known up to 27~keV from
\RXTE\ data \citep{kuiperhermsen15},
and improved its spectral characterization at these hard X-ray energies.
For the other young neutron star \cxo\ in the supernova remnant G18.95$-$1.1,
we better determined its powerlaw spectrum with the new \Chandra\ data.

For the two old millisecond pulsars, \psreight\ and \psrnine, \Chandra\
HRC images yielded their detections in X-ray for the first time.
Unfortunately, they are both relatively faint, and significantly better
information on them in X-ray may only be achieved with much longer exposures.
Nonetheless, \psreight\ and \psrnine\ are worth further study.
These two belong to a small group of binary pulsars
with a spin period $P<6\mbox{ ms}$
and a $\gtrsim0.3\Msun$ carbon-oxygen white dwarf companion;
others include
PSR~J1101$-$6424 \citep{ngetal15},
PSR~J1125$-$6014 \citep{reardonetal21,shamohammadietal23},
PSR~J1614$-$2230 \citep{demorestetal10,alametal21},
PSR~J1618$-$4624 \citep{cameronetal20}, and
PSR~J1943+2210 \citep{scholzetal15}.
These systems are thought to be descendants of intermediate-mass X-ray
binaries \citep{taurisetal11,taurisetal12}.
Prior to the X-ray detections presented here,
only two had been detected outside of radio:
PSR~J1125$-$6014 in gamma-rays \citep{smithetal19}
and PSR~J1614$-$2230 at gamma-ray and X-ray energies
\citep{abdoetal10,marellietal11,pancrazietal12}.
Their high-energy detections are probably due in part to them being
the closest of the seven named above, in particular, PSR~J1614$-$2230
is at a distance of $670^{+50}_{-40}\mbox{ pc}$ \citep{arzoumanianetal18} and
PSR~J1125$-$6014 is at $1.4\pm0.3\mbox{ kpc}$ \citep{reardonetal21}.
In contrast, \psrnine\ is at $1.6^{+0.2}_{-0.3}\mbox{ kpc}$ \citep{geyeretal23},
\psreight\ is at a radio DM inferred distance of 0.2 or 1~kpc
(see \citealt{zhangetal26} and Section~\ref{sec:intro}),
and the other three are at DM distances of $\sim2$, 4, and 6~kpc, respectively.

It is also noteworthy that
some millisecond pulsars with high spin-down luminosity $\dot{E}$ are
bright in X-rays due to heating of their polar caps by return currents
(see, e.g., \citealt{arons81,hardingmuslimov01,hardingmuslimov02}).
PSR~J1614$-$2230 and PSR~J1125$-$6014 have the highest $\dot{E}$ of
the above seven at $1.2\times10^{34}\mbox{ erg s$^{-1}$}$ and
$8\times10^{33}\mbox{ erg s$^{-1}$}$, respectively.
On the other hand, \psrnine\ and \psreight\ have
$\dot{E}=2.1\times10^{33}\mbox{ erg s$^{-1}$}$ and
$\dot{E}=9.5\times10^{32}\mbox{ erg s$^{-1}$}$, respectively,
and the remaining three have
$\dot{E}=(2.6,0.6,0.5)\times10^{33}\mbox{ erg s$^{-1}$}$.
Taken together, especially its relative distance, the X-ray brightness
of \psrnine\ is perhaps unsurprising.

In comparison, the X-ray flux of \psreight\ is about 1.5 times lower than
that of \psrnine,
such that if it has the same X-ray luminosity (and same as PSR~J1614$-$2230),
then the distance to \psreight\ is about 2~kpc, which is much further than
implied by its DM.  On the other hand, the $\dot{E}$ of \psreight\ is more
than ten times lower than that of PSR~1614$-$2230.
A distance of 1~kpc is still possible if its intrinsic X-ray luminosity
is four times lower,
while a distance greater than 2~kpc is possible if its luminosity is greater
than that of PSR~J1614$-$2230 and \psrnine\ despite its low DM and $\dot{E}$.
As a consequence, we suggest 2~kpc as an estimate of the upper bound on the
distance, which means \psreight\ is still one of the closer systems in its group
and could merit future study, especially since future radio monitoring
could yield a measurement of a high neutron star mass \citep{zhangetal26}.

\section*{Acknowledgements}

WCGH thanks Di Li and Lei Zhang for discussions and
Patrick Slane for approving the Director's Discretionary Time (DDT)
observation of \psreight.
PHW thanks Timothy Leon for his guidance on spectral extraction and analysis.
The authors thank the anonymous referee for comments that led to improvements
in the paper.
WCGH acknowledges support through grant 80NSSC26K0369 from NASA
and Chandra award SAO GO6-27040X.
Chandra grants are issued by the Chandra X-ray Center (CXC),
which is operated by the Smithsonian Astrophysical Observatory
for and on behalf of NASA under contract NAS8-03060.
S.G. is supported by the CNES and by the ANR (Agence Nationale de la
Recherche) grant ANR-25-CE31-7901-01 (ANR DENSER)

This research made use of data obtained from the Chandra Data
Archive and the Chandra Source Catalog, and software provided by the
Chandra X-ray Center (CXC) in the application packages CIAO and Sherpa.
This paper employs a list of Chandra datasets, obtained by the Chandra
X-ray Observatory, contained in the Chandra Data Collection (CDC)~683
(\href{https://doi.org/10.25574/cdc.683}{doi:10.25574/cdc.683}).
This work is based on observations obtained with \XMM, an ESA
science mission with instruments and contributions directly funded
by ESA Member States and NASA.
This work uses data and software provided
by the High Energy Astrophysics Science Archive Research Center
(HEASARC), which is a service of the Astrophysics Science Division
at NASA/GSFC and High Energy Astrophysics Division of the
Smithsonian Astrophysical Observatory.
This research made use of data from the \NuSTAR\ mission, a project led by
the California Institute of Technology, managed by the Jet Propulsion
Laboratory, and funded by NASA. Data analysis was performed using
the NuSTAR Data Analysis Software (NuSTARDAS), jointly developed by the
ASI Science Data Center (SSDC, Italy) and the California Institute of
Technology (USA).
This work made extensive use of the NASA Astrophysics Data System
(ADS) Bibliographic Services and the arXiv.

%%%%%%%%%%%%%%%%%%%%%%%%%%%%%%%%%%%%%%%%%%%%%%%%%%
\section*{Data Availability}

Data underlying this article will be shared on reasonable request
to the corresponding author.

%%%%%%%%%%%%%%%%%%%% REFERENCES %%%%%%%%%%%%%%%%%%

\bibliographystyle{mnras}
\bibliography{xns26}

%%%%%%%%%%%%%%%%%%%%%%%%%%%%%%%%%%%%%%%%%%%%%%%%%%

\bsp
\label{lastpage}
\end{document}